\documentclass[10pt,journal,compsoc]{IEEEtran}
\newif\ifpeerreview

\peerreviewfalse

\usepackage{url}
\usepackage{amsmath,amssymb,graphicx}

\usepackage{lipsum} 

\usepackage[switch]{lineno}

\usepackage{cite}
\usepackage{url}
\usepackage{algorithm,algorithmic,multicol}
\usepackage{xcolor} 
\usepackage{amsmath}
\usepackage{amssymb}
\let\conjugatet\overline
\DeclareMathAlphabet{\pazocal}{OMS}{zplm}{m}{n}
\DeclareMathOperator{\E}{\mathbb{E}}
\usepackage{graphicx}

\newcommand{\paperID}{48}

\title{Adaptive Gradient Balancing for Undersampled MRI Reconstruction and Image-to-Image Translation}

\author{Itzik Malkiel, Sangtae Ahn, Valentina Taviani, Anne Menini, Lior Wolf, Christopher J. Hardy

\IEEEcompsocitemizethanks{
\IEEEcompsocthanksitem I. Malkiel and L. Wolf are with School of Computer Science, Tel Aviv University, Israel.

\IEEEcompsocthanksitem S. Ahn and C. Hardy are with GE Research, Niskayuna, NY,
USA.
\IEEEcompsocthanksitem V. Taviani and A. Menini are with GE Healthcare, Menlo Park, CA, USA.
}
}

\begin{document}

\IEEEtitleabstractindextext{%
\begin{abstract}

Recent accelerated MRI reconstruction models have used Deep Neural Networks (DNNs) to reconstruct relatively high-quality images from highly undersampled $k$-space data, enabling much faster MRI scanning. However, these techniques sometimes struggle to reconstruct sharp images that preserve fine detail while maintaining a natural appearance. 
In this work, we enhance the image quality by using a Conditional Wasserstein Generative Adversarial Network combined with a novel Adaptive Gradient Balancing (AGB) technique that automates the process of combining the adversarial and pixel-wise terms and streamlines hyperparameter tuning. In addition, we introduce a Densely Connected Iterative Network, which is an undersampled MRI reconstruction network that utilizes dense connections. In MRI, our method minimizes artifacts, while maintaining a high-quality reconstruction that produces sharper images than other techniques. To demonstrate the general nature of our method, it is further evaluated on a battery of image-to-image translation experiments, demonstrating an ability to recover from sub-optimal weighting in multi-term adversarial training.
\end{abstract}

}

\maketitle
\thispagestyle{empty}

\begin{figure*}[h]
\includegraphics[width=0.85\linewidth]{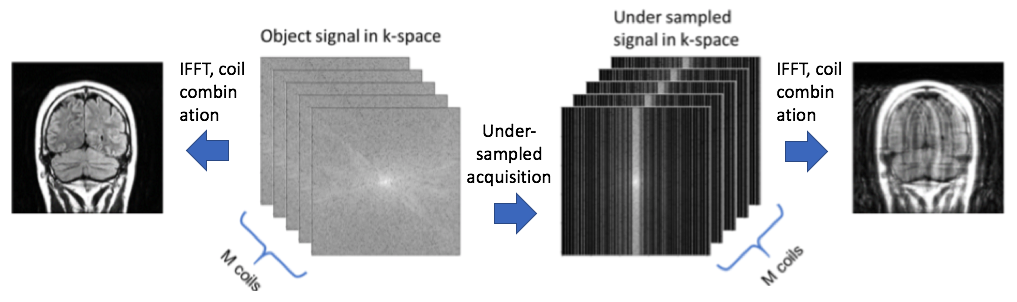}
  \centering
\vspace{-8pt}
    \caption{Fully-sampled $k$-space multiplied by an acquisition sampling pattern, with acceleration factor of 4, results in highly undersampled $k$-space. Reconstruction of the undersampled $k$-space using zero-filling generates a low-quality image with heavy artifacts that is completely non-diagnostic (right). A non-accelerated acquisition that uses fully-sampled $k$-space results in high-quality image (left). In this study, we focus on 2D data acquisition, which utilizes a 1D sampling pattern in the phase-encoding direction.
  }
  \label{fig:introFig}
    
\end{figure*}

\IEEEraisesectionheading{
  \section{Introduction}\label{sec:introduction}
}
\vspace{-7pt}
\IEEEPARstart{M}{agnetic} resonance imaging (MRI) data acquisition is inherently slow, often exceeding 30 min. per exam. 
One way to accelerate MR scanning is by undersampling $k$-space, i.e., reducing the number of $k$-space traversals by a factor $R$, accelerating the scan proportionately. However, this violates the Nyquist criterion, resulting in aliasing artifacts in the zero-filled reconstructed image (Fig.~\ref{fig:introFig}). Over the last several decades, a number of methods have been used to overcome this problem, including Parallel Imaging (PI) \cite{1sodickson1997simultaneous, 2pruessmann1999sense, 3griswold2002generalized} and Compressed Sensing (CS) \cite{5lustig2007sparse}. 

More recently, Deep Neural Networks (DNNs) have been used to push $R$ values even higher \cite{8hammernik2018learning,9schlemper2018deep,6zhu2018image}. Among the most promising Deep Learning (DL) techniques, unrolled iterative networks (also called cascading networks) have emerged as a leading powerful method~\cite{8hammernik2018learning,9schlemper2018deep}. Inspired by CS, this technique uses a DNN composed of a sequence of iterations that include data-consistency and regularization units. The data-consistency units utilize the acquired $k$-space lines as a prior that keeps the network from drifting away from the acquired data, and the regularization units are trained to regularize the reconstruction.

As with other image generation problems, using a naive pixel-wise distance for training DL-based undersampled MRI reconstruction models can result in image blurring and unrealistic appearance. In a clinical setting, avoidance of blurring can be crucial for proper diagnosis. Recently, Generative Adversarial Networks (GANs) have been used to promote the naturalness of MRI reconstructions\cite{14hammernik2018variational,12mardani2019deep,13yang2018dagan}. In our work, we harness the power of conditional Wasserstein GANs (cWGANs) to further improve image quality, and alleviate long training and experimental process using a novel training technique for multi-term adversarial objectives.

The main contributions of this paper are as follows: (1) We 
{adopt and evaluate} a cWGAN method for undersampled MRI reconstruction, in which both the generator and discriminator are conditioned using the acquired undersampled data. (2) We introduce a novel training algorithm called Adaptive Gradient Balancing (AGB) which balances the losses in multi-term adversarial objectives. 
(3) We provide an extensive comparison between different models and training techniques. In particular, we report results of six methods---an unrolled iterative network, a Variational Network\cite{8hammernik2018learning,14hammernik2018variational}, a CNN-Cascade \cite{9schlemper2018deep}, a WGAN based network, a cWGAN, and a cWGAN trained with our AGB. 
(4) We propose and evaluate a novel Densely Connected Iterative Network (DCI-Net) for undersampled MRI reconstruction, which is inspired by Dense-Nets \cite{10huang2017densely}. 

\vspace{-12pt}
\section{Related work}

Recent methods in image-to-image translation adopt the idea of conditional GANs (cGANs) \cite{17mirza2014conditional}, in which the generated data $g$ are conditioned by data $y$ that are being fed to both the generator and discriminator networks. Isola et al. \cite{20isola2017image} uses a cGAN \cite{17mirza2014conditional} to learn a mapping from one image domain to another, such as converting a satellite photo into a map, a sketch into a photorealistic image, etc. Wang \cite{19wang2018high} extended this work to generating high-resolution images.

There have recently appeared an increasing number of DL-based accelerated MRI reconstruction models {\color{black}\cite{9schlemper2018deep,15diamond2017unrolled,malkiel2018densely, malkiel2019leveraging,8hammernik2018learning,14hammernik2018variational,13yang2018dagan, hardy2018residual, brada2019towards, rotman2021correcting, chen2018variable,chen2018improving}}
. Schlemper et al. \cite{9schlemper2018deep} used a cascade of convolutional neural networks (CNNs) employing data-consistency layers and optimized to minimize a pixel-wise distance. Diamond et al. \cite{15diamond2017unrolled} proposed a framework for integrating prior knowledge into DL architecture called unrolled optimization with deep priors (ODP). They presented a general method for solving inverse imaging problems and demonstrated their approach on undersampled MRI reconstruction. Zhu et al. \cite{6zhu2018image} introduced automated transform by manifold approximation (AUTOMAP), a DNN that learns a mapping between sensor and image domains. The AUTOMAP architecture is composed of fully connected layers followed by convolutional layers trained to optimize a pixel-wise objective. 

Hammernik et al. \cite{8hammernik2018learning,14hammernik2018variational} proposed a variational network (VN) for solving undersampled MRI reconstruction. In \cite{8hammernik2018learning} they presented a VN trained to minimize a pixel-wise loss, and then in \cite{14hammernik2018variational} proposed a GAN-based VN to reduce blurring and to improve the perceptual appearance of the reconstructed images. Mardani et al. \cite{12mardani2019deep} proposed a GAN-based model that uses a deep residual network as a generator, and a discriminator trained to optimize a mixture of pixel-wise and least squares GAN (LSGAN) losses \cite{16mao2017least}. 
In \cite{yu2017deep,13yang2018dagan}, deep de-aliasing GAN (DAGAN) was introduced, a GAN-based model trained to optimize a mixture of pixel-wise, perceptual and GAN losses. In contrast to the original cGAN technique \cite{17mirza2014conditional}, DAGAN uses an architecture that conditions only the generator input but not the discriminator, which in this context is similar to \cite{12mardani2019deep,14hammernik2018variational}. The authors also reported \cite{13yang2018dagan} that a model that uses only pixel-wise and GAN (but not perceptual) losses, generates unrealistic jagged artifacts. As a motivation to our study, we experienced similar behavior when training our generator to minimize a weighted-sum objective with a weighting that appeared to favor the GAN term.

Our method differs from other DL based undersampled reconstruction studies in employing a conditional architecture that conditions both the generator and the discriminator, as we found that applying a conditional discriminator has a profound impact on model convergence and performance. In addition, our method is unique in its AGB training and DCI-Net generator architecture. The former improves performance, {automates the process of multi-term adversarial loss combination, }
and streamlines hyperparameter tuning. The latter provides a simple yet effective technique for promoting feature propagation and reuse in iterative networks.

{\color{black}Very recently, the authors of \cite{pnas2020instabilities} have reported that DL-based reconstruction methods can suffer from instabilities. Our study was motivated by similar concerns.}

{\color{black}In \cite{lucic2017gans}, the authors show that most GAN techniques, including \cite{29arjovsky2017wasserstein,gulrajani2017improved, mao2016least, berthelot2017began, kodali2017convergence}, can reach similar scores with sufficient exploration of hyper-parameters and random initializations. In addition, the authors show that most improvements do not arise from fundamental algorithmic changes of the underlying GAN technique. In our work, we propose a technique that can accelerate the process of hyperparameter tuning.} 

\begin{figure*}[t]
\includegraphics[width=0.7305\linewidth]{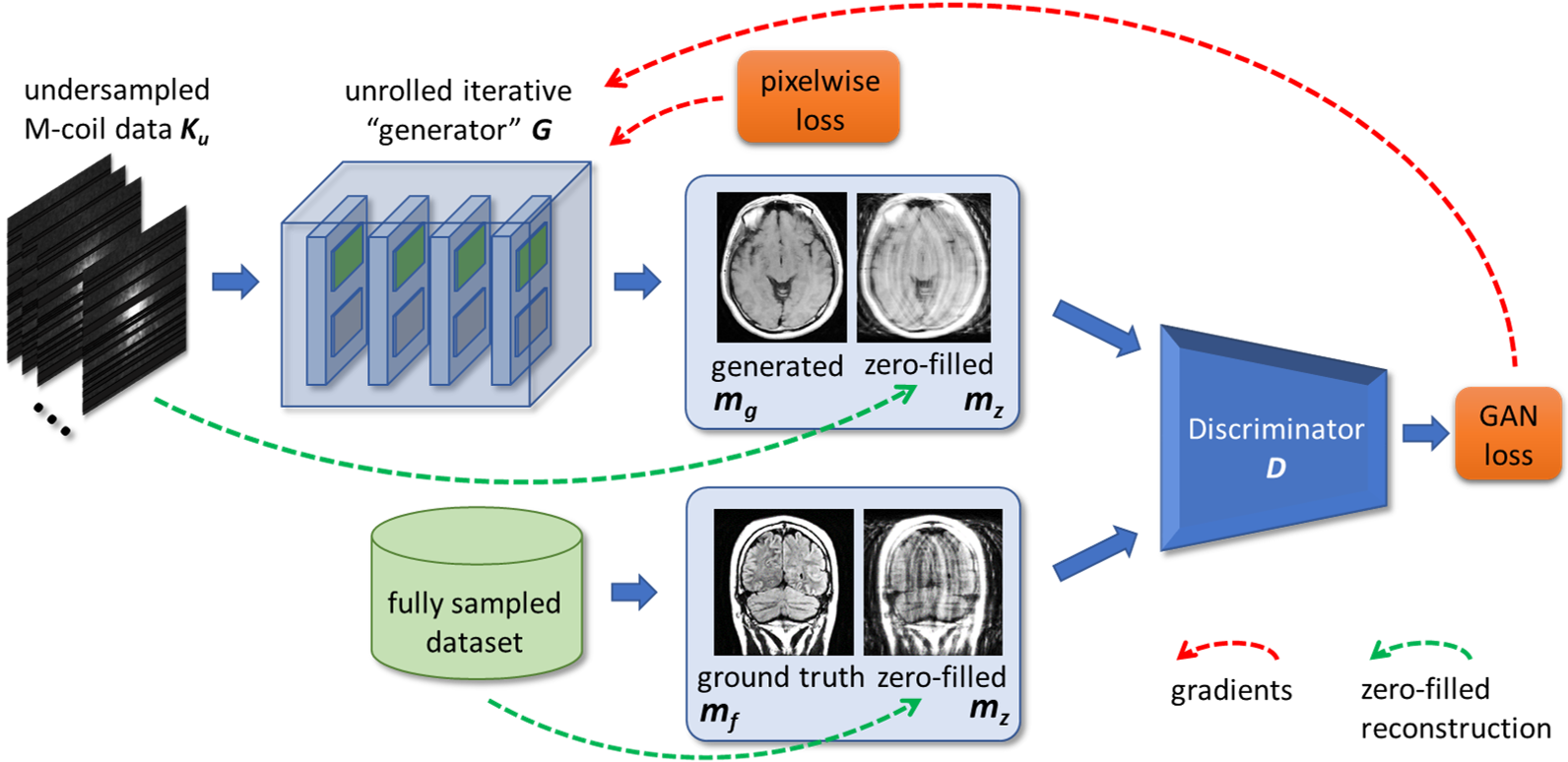}
\centering
\caption{The generator receives undersampled $k$-space data as input and generates a matched estimated fully-sampled image. The discriminator learns to estimate the Wasserstein Distance between “fake” pairs and “real” pairs. 
}
\label{fig:cgan}
\end{figure*}

\vspace{-6pt}
\section{Problem Formulation}
Let $k \in \mathbb{C}^{H \times W}$ be the $k$-space signal acquired by an MRI scanner. W and H are the width and height of the acquired signal. For a single-coil receiver, an image $m \in \mathbb{C}^{H \times W}$ can be estimated by performing an inverse Fourier transform:
$m=\mathcal{F}^{-1}(k).$
In multi-coil MRI, an array of $N$ coils acquire $N$ different 2D $k$-space measurements of the same object
\begin{equation}
K = \left\{ k_i | k_i \in \mathbb{C}^{H \times W}, i = 1\dots N \right\}.
\end{equation}
Each coil $C_i$, positioned at a different location, is typically highly sensitive in one region of space. This position-dependent sensitivity can be represented by a complex-valued coil sensitivity map in real space, 
\begin{equation}
S = \left\{ s_i | s_i \in \mathbb{C}^{H \times W}, i = 1\dots N \right\}.
\end{equation}

During reconstruction, the images from each coil are combined into a fully-sampled image 
\begin{equation}
m_f = \pazocal{R}\left(K,S\right)
\end{equation}
where $\pazocal{R}$ is a reconstruction function
\begin{equation}
\pazocal{R}\left(K,S\right) = \sum\limits_{i=1}^N \conjugatet{s_i} \odot \mathcal{F}^{-1}(k_i).
\end{equation}
$\conjugatet{s_i}$ is the complex conjugate of the sensitivity map of coil $C_i$, and $\odot$ denotes the Hadamard product. To accelerate imaging, a binary sampling pattern $M$ is used to undersample each coil's $k$-space signal for each slice. The undersampled $k$-space signal, denoted by $K_u$, can be calculated by 
\begin{equation}
K_u = M \odot K.
\end{equation}
The undersampled zero-filled image $m_z$ can be written as
\begin{equation}
m_z = \pazocal{R}\left(K_u, S\right).
\end{equation}
The learning task is to find a reconstruction function $G^{*}$ that minimizes an expected loss function $\pazocal{L}$ (Sec.~\ref{sec:Objective}) over a population of scans: 
\begin{equation}
G^{*} = \arg\min_{G}\mathbb{E}_{\left(K,M\right)}\left[\pazocal{L}\left(G \left( K_u\right)\right) \right].
\end{equation}
For a given $G^*$ and $K_u$, we will denote the generated image by 
\begin{equation}
m_g := G^{*}\left(K_u\right).
\end{equation}

\vspace{-11pt}
\section{Methods}

Our method learns a DL-based undersampled MRI reconstruction model from training samples, each of which is a pair of a fully sampled and matched undersampled $k$-space data. We propose a cGAN architecture, which conditions the reconstruction using the  zero-filled image. Specifically, our model is composed of a generator and a discriminator network. The generator reconstructs an image from undersampled $k$-space data. The discriminator receives a pair of input images: (i) either a ground truth image $m_f$ or a generated (``fake'') image $m_g$ from undersampled $k$-space data and (ii) a zero-filled image $m_z$ (see Fig.~\ref{fig:cgan}).

While it is possible to use a non-conditional GAN architecture that looks only at $m_g$ or $m_f$, in this case the discriminator can only enforce general style properties learned from the distribution of fully sampled images, which might not necessarily match a specific image. For example, a discriminator that learns that fully sampled images possess a high sharpness level may encourage all generated images to maintain sharp features, although some features might be less sharp than others. Under the context of training, such a non-conditional discriminator can be fooled by fake images that are inconsistent with their prior information from $m_z$, and thus can generate gradients that do not necessarily correlate with a pixelwise term. This property can hinder training convergence (see Sec.~\ref{sec:convergance}), and degrade data fidelity, and may expose the generator to hallucinations. However, a conditional discriminator can enforce both realistic appearance and spatial consistency by matching each generated image with its corresponding zero-filled image.

\vspace{-8pt}
\subsection{Objective} 
\label{sec:Objective}
Following the success of the Wasserstein GAN (WGAN)~\cite{29arjovsky2017wasserstein} and the framework proposed by Isola et al.~\cite{20isola2017image}, we adopt a conditional WGAN objective:
{\color{black}
\begin{multline}
\pazocal{L}_{cWGAN}(G,D) = \E_{(m_z, m_f)}[D(m_z,m_f)] - \\ \E_{(m_z, K_u)}[D(m_z,G(K_u))] 
\end{multline}
where $G$ and $D$ are the generator and discriminator networks, repectively. $K_u$ is an undersampled $k$-space dataset, $m_f$ is a fully sampled image and $m_z$ is a zero-filled image.} In addition to the adversarial loss, we also add a pixel-wise Mean Square Error (MSE) loss  
\begin{equation}
\pazocal{L}_{MSE}(G)= \frac{1}{WH}\sum_{i=1}^{W}\sum_{j=1}^{H}((m_f)_{i,j}-G(K_u)_{i,j})^2.
\end{equation}
can be expressed as: 
\begin{equation}
\pazocal{L}(G) = \max_D \frac{1}{\beta} \pazocal{L}_{cWGAN}(G,D) +\pazocal{L}_{MSE}(G)
\end{equation}
where $\beta$ is an adaptive weight that changes during training (see next section).

\vspace{-8pt}
\subsection{Adaptive Gradient Balancing}
In WGAN training, the discriminator network is used as a learned loss function, which dynamically changes during training, and thus may generate gradients with variable norm. To stabilize the WGAN training and to avoid drifting away from the ground-truth spatial information, we introduce the Adaptive Gradient Balancing (AGB) algorithm for continually balancing the gradients of the pixel-wise and WGAN loss functions. 

In order to keep the gradients of both terms at the same level, and since the WGAN gradients tend to vary, we choose to adaptively upper-bound the WGAN gradients. Specifically, we define $\beta$ to be an adaptive weight that will be used to bound the WGAN loss gradients. We calculate two moving-average variables $g_{ma}$ and $p_{ma}$ corresponding to the WGAN loss and the pixel-wise loss, respectively. These moving averages capture the standard deviation (SD) of the gradients calculated at every backward step on the generated image, with respect to each one of the losses separately. At every training step, if
$g_{ma} > p_{ma} \cdot ratio$
for a predefined $ratio$ value, we update $g_{ma}$ and $\beta$ as follows:
$\beta \gets \beta \cdot (1 + rate)$,
$g_{ma} \gets g_{ma} \cdot (1 - rate)$,
where $rate$ is a predefined decay rate. During training, we divide the WGAN loss by $\beta$ to carefully decay the WGAN loss gradients to roughly the same order of magnitude as those of the pixel-wise loss. Moreover, in order to keep a reasonable ratio between the generator's WGAN loss gradients and the discriminator loss gradients, we also decay the discriminator loss by the same $\beta$ factor (see Alg.~1).

\begin{figure*}[t]
\includegraphics[width=0.7\linewidth]{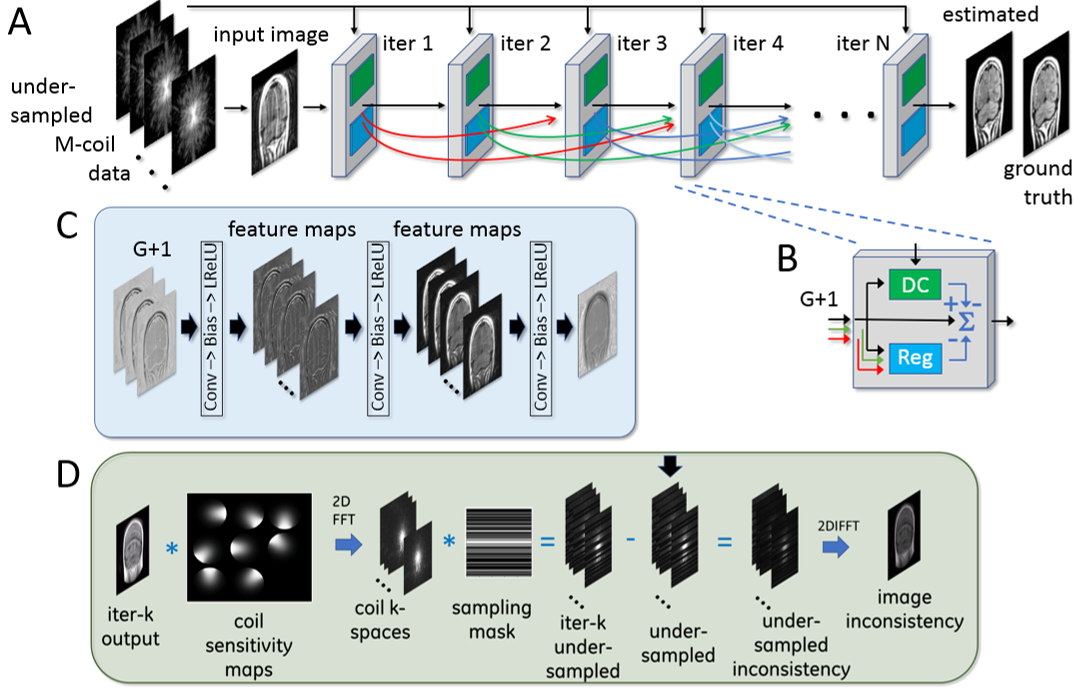}
  \centering
  \caption{DCI-Net (A) consists of N unrolled iterative blocks, each with dense skip-layer connections (curved arrows) to subsequent blocks. Each iterative block (B) consists of data-consistency (D) and regularization (C) units. The regularization unit operates on all G+1 connections, while the data-consistency unit operates only on direct connection. Images at all stages are complex---in practice, treated by creating real and imaginary channels (not shown). 
  }
    \label{fig:dci}
\end{figure*}

\begin{algorithm*}[t] 
  \caption{AGB training of WGANs for multi-term loss. Here $\alpha = 5 \cdot 10^{-5}$, $\beta_{init} = 10$, $c = 0.01$, $\lambda = 0.99$, $ratio = 10$, $rate = 0.01$, $n_{discriminator} = 1$. {\color{black}$D_w$ and $G_\theta$ are discriminator and generator networks with weights $w$ and $\theta$, respectively.}}
  \begin{minipage}[t]{0.35\textwidth}
  \scriptsize
  {\color{black}
  \hspace{0cm}
  {
  $p_{ma} \gets 0$; $g_{ma} \gets 0$; $\beta \gets \beta_{init}$ }\\
  \hspace*{0cm}
  {
  \textbf{for} number of training iterations \textbf{do} }\\
  \hspace*{.35cm}
  {
  \textbf{for} $t$ = 0, ..., $n_{discriminator}$ \textbf{do} }\\
  \hspace*{.7cm}
  {
  Sample a minibatch \{($K_u^{i}, m_z^{i}, m_f^i$)\}$_{i=1}^{m}$ }\\
  \hspace*{.7cm}
  {
  $g_w$ $\gets$ $\nabla _w$ $[\frac{1}{m}\frac{1}{\beta}$ $(\sum_{i=1}^{m} D_w(m_z^{i}, m_f^{i}) - $ }\\
  \hspace*{1.05cm}
  {
  $\sum_{i=1}^{m} D_w(m_z^{i}, G_\theta(K_u^{i})))]$  }\\
  \hspace*{.7cm}    
  {
  $w$ $\gets$ $w$ + $\alpha$ $\cdot$ Adam($w$, $g_w$) }\\
  \hspace*{.7cm}
  {
  $w$ $\gets$ clip($w$, -$c$, $c$) }\\
  \hspace*{.35cm}
  {
  \textbf{end for} }}
  \end{minipage}
  \begin{minipage}[t]{0.35\textwidth}
  \scriptsize
  {\color{black}
  \hspace*{0cm}
  {
  Sample a minibatch \{($K_u^{i}, m_z^{i}, m_f^i$)\}$_{i=1}^{m}$ }\\
  \hspace*{0cm}
  {
  $g_\theta$ $\gets$  $  \nabla _\theta$ $ [$ $\frac{1}{m}(- \frac{1}{\beta}\sum_{i=1}^{m} D_w(m_z^{i}, G_\theta(K_u^{i}))$ }\\
  \hspace*{0.35cm}
  {
  $ +\sum_{i=1}^{m} MSE(m_f^i, G_\theta(K_u^{i})))]$ }\\
  \hspace*{0cm}        
  {
        $\theta$ $\gets$ $\theta + \alpha$ $\cdot$ Adam($\theta$, $g_\theta$) }\\
  \hspace*{0cm}
  {
       $g_{gan}$ $\gets$ $\frac{1}{m}\frac{1}{\beta}\sum_{i=1}^{m}  \nabla _{G_\theta(K_u^{i})} [D_w(m_z^{i}, G_\theta(K_u^{i}))]$ }\\
  \hspace*{0cm}
  {
       $g_{MSE}$ $\gets$ $ \frac{1}{m}\sum_{i=1}^{m}  \nabla _{G_\theta(K_u^{i})} [MSE(m_f^i, G_\theta(K_u^{i}))]$ }\\
  \hspace*{0cm}
  {
       $g_{ma} \gets g_{ma} \cdot \lambda + (1-\lambda) \cdot SD(g_{gan})$ }\\
  \hspace*{0cm}
  {
       $p_{ma} \gets p_{ma} \cdot \lambda + (1-\lambda) \cdot SD(g_{MSE})$ }\\
    }
  \end{minipage}
  \begin{minipage}[t]{0.3\textwidth}
  \scriptsize
  {\color{black}
  \hspace{0.85cm}
  {
  \textbf{if} $g_{ma} > p_{ma} \cdot ratio $ \textbf{then} }\\
  \hspace*{1.2cm}
  {
  $\beta \gets \beta \cdot (1 + rate)$ }\\
  \hspace*{1.2cm}
  {
  $g_{ma} \gets g_{ma} \cdot (1 - rate)$ }\\
  \hspace*{0.85cm}
  {
  \textbf{end if} }\\
  \hspace*{0.5cm}
  {
  \textbf{end for} }
  }
  \end{minipage}
  
\end{algorithm*}

By extending WGAN training to adaptively balance a multi-term loss objective, our AGB algorithm ensures one invariant during the entire training---the SD of the WGAN loss gradients is upper-bounded by a factor of the SD of the pixel-wise gradients. This invariant maintains the effectiveness of both loss terms, over the course of training.  

\subsubsection{Hyperparameteres}
AGB utilizes the following hyperparameters: $\beta_{init}$ (a value for initializing the decay rate for the GAN gradients), $\lambda$ (decay weight used to calculate the moving average of the gradients of each loss), $ratio$ (defines the maximal ratio between the GAN loss gradients and the pixel-loss gradients), $rate$ (the rate used to increase the $\beta$ value), learning rate, and clipping value (for the WGAN training). {\color{black} The same default parameters are used in all experiments. $\beta_{init} = 10$, $\lambda = 0.99$, $ratio = 10$, $rate = 0.01$.}

The $ratio$ is the only parameter for which we applied hyperparameter search during the development of the method (besides the early research phase of developing the method, we did not employ any hyperparameter search in any of the experiments). We have explored the values 1, 10, 100, 1000, and found that configuring $ratio$ with a value of 10, which upper bounds the GAN gradients by a factor of 10 compared to the pixel-loss gradients, yields the best performance. Then, through all of our experiments, we set $ratio = 10$.

In our experiments, the learning rate was chosen to be the same as the one used in the original pix2pix work~\cite{20isola2017image} or the value used during the training of the MRI reconstruction network when optimizing only the pixel-wise loss. The clipping value is the recommended one from WGAN.

\begin{figure*}[t]
\centering
\begin{tabular}{ccc}
    \includegraphics[width=.3\linewidth]{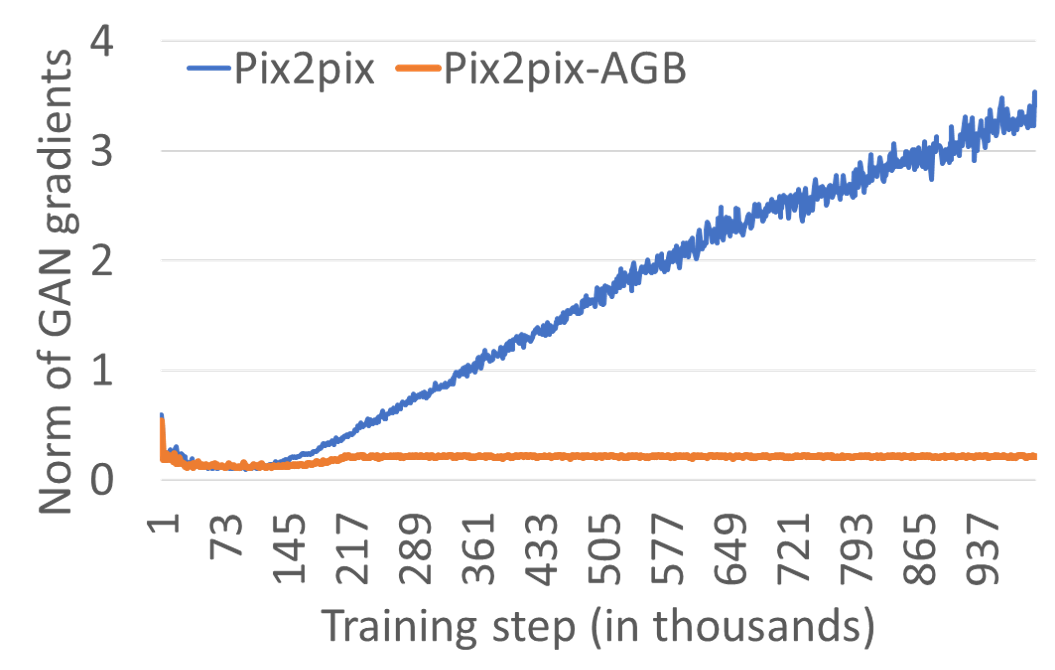}&
    \includegraphics[width=.3\linewidth]{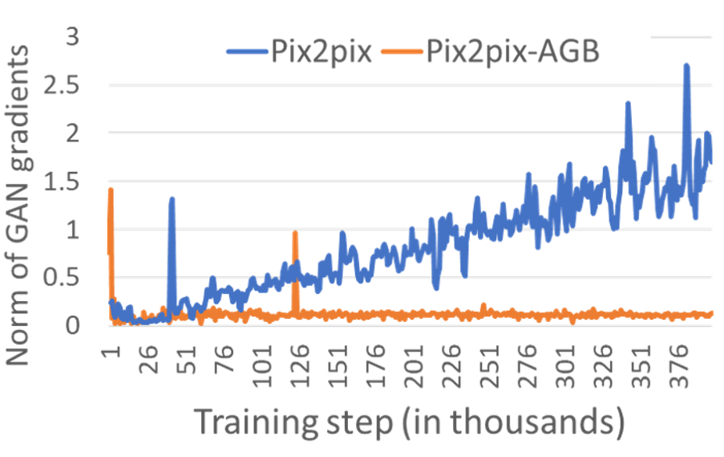}&
    \includegraphics[width=.3\linewidth]{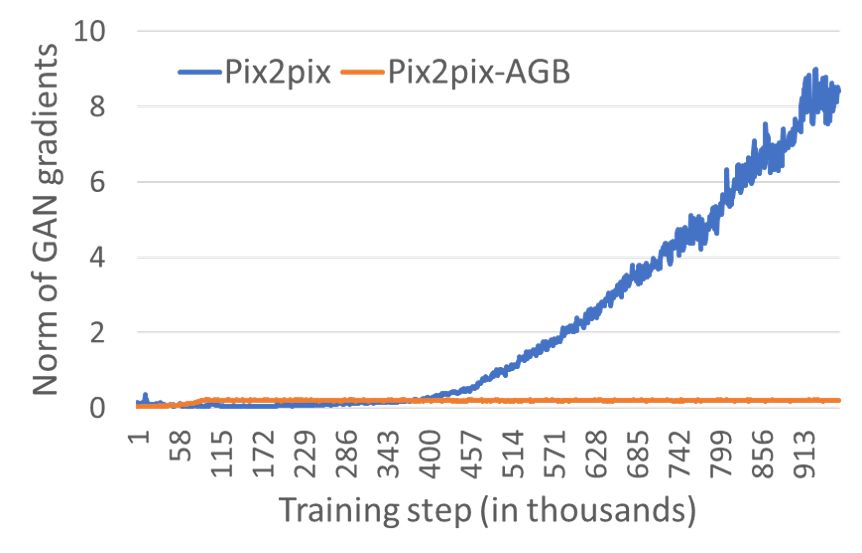}    \\
    (a) & (b) & (c) \\
    \end{tabular}
  \caption{{\color{black}Moving averages calculated on the norm of the GAN loss gradients, versus training step, for Pix2pix and Pix2pix-AGB trained to generate (a) facade images from annotations, (b) urban street scenes from annotations, and (c) aerial photographs from map images. The gradients are derived with respect to the pixels of the generated image. The increasing norm may indicate the growing dominance of the GAN loss during training. The norm of the pixel-wise loss gradients (not shown in the figure) remains fixed since Pix2pix utilizes an L1 loss. Both models were trained for 1000 epochs (instead of 200, as suggested in the Pix2pix work).}
    }
    \label{fig:pix2pix_gradients}
\end{figure*}

\vspace{-8pt}
\subsection{Network Architectures}
\subsubsection{Generator}
We propose a new generator architecture (Fig.~\ref{fig:dci}), called Densely Connected Iterative Network (DCI-Net), based on the iterative CNN \cite{8hammernik2018learning,9schlemper2018deep}. 
The key new developments are the use of (1) dense connections~\cite{10huang2017densely} across all iterations, which strengthens feature propagation, making the network more robust, and (2) a relatively deep architecture of over 60 convolutional layers, bringing increased capacity.  Our generator receives M coils of undersampled $k$-space data, and uses N = 20 iterations, each of which includes a data-consistency unit and a regularization unit (Fig.~\ref{fig:dci}B). Dense skip-layer connections between the output of each iteration and the following G iterations---where typically G = 5---are represented as curved lines in Fig.~\ref{fig:dci}A. This results in an input to each block composed of skip and direct connections concatenated to form a G+1 channel complex image.

\textbf{Data-consistency unit}
Each data-consistency (DC) unit (Fig. 3D) shades the input image with each coil sensitivity map, transforms the resulting images to $k$-space, imposes the sampling mask, calculates the difference relative to acquired $k$-space and returns them to the image domain, multiplied by a learned weight (see Fig.~\ref{fig:dci}D). {\color{black}These operations can also be expressed as:
\begin{equation}
DC(m_f^k) =  \pazocal{R} (\mathcal{F} (m_f^k \odot S) \odot M - K_u, S) \cdot \lambda_k
\end{equation}
where $m_f^k$ and $\lambda_k$ are the input image and the learnt weight of the $k$th iteration, respectively.} By utilizing the acquired $k$-space data as a prior, the data-consistency units, embedded as operations inside the network, keep the network from drifting away from the acquired data. For this use, the undersampled $k$-space data were also input directly into each iterative block of the network (Figs.~\ref{fig:dci}A,B).

\begin{figure}[t]
  \includegraphics[width=1.0\linewidth]{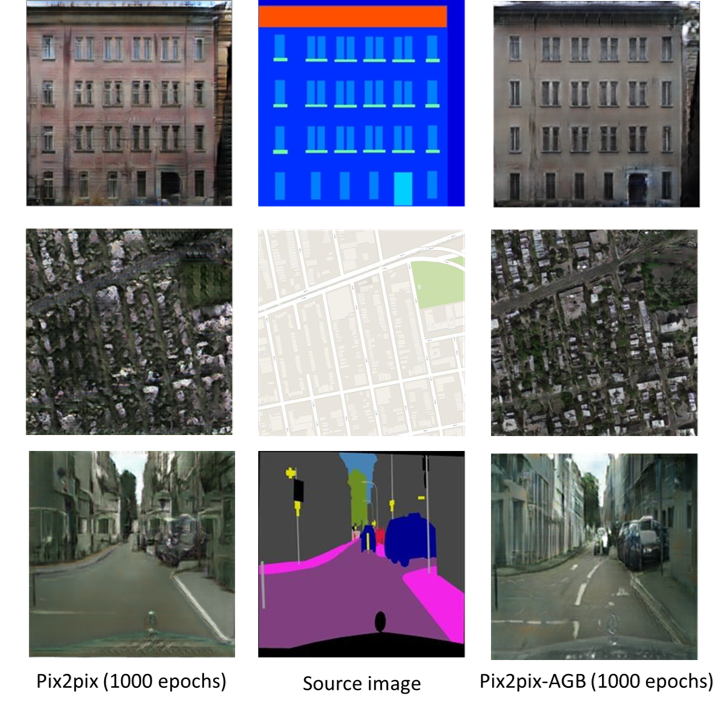}
  \centering
  \caption{{\color{black}Representative samples from the Facade, cityscapes and maps test sets. The same source image (center) was used by both models as a source to generate the images on either side. Pix2pix with a longer training of 1000 epochs (left) introduces visual artifacts. Pix2pix-AGB trained for the same number of epochs (right) yields higher image quality.}
  }
    \label{fig:Quali_pix2pix_1000}
\end{figure}

\begin{figure}[t]
  \includegraphics[width=0.95\linewidth]{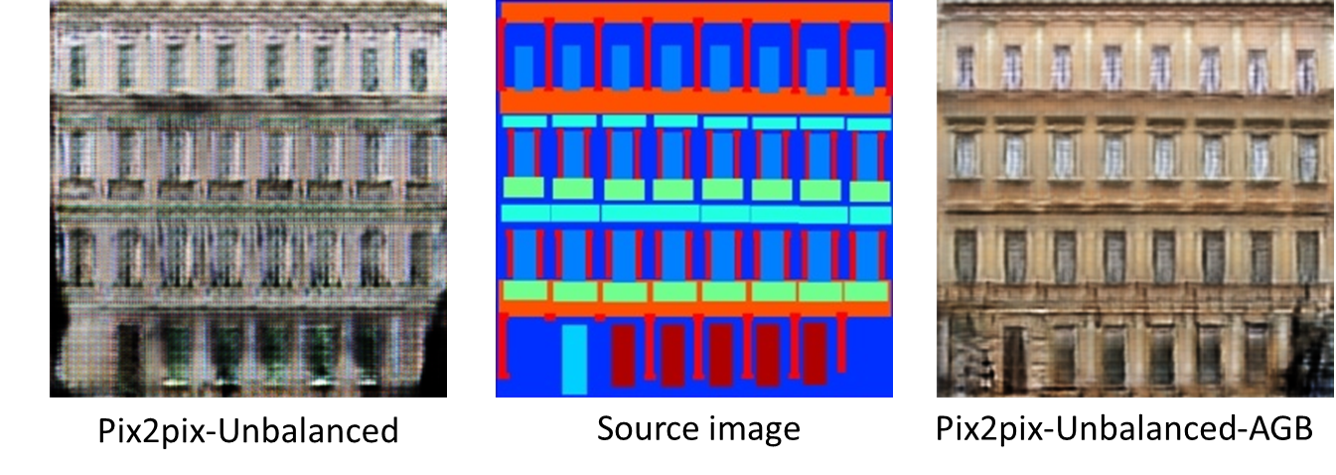}
  \centering
  \caption{A representative sample from the Facade test set. The same facade annotation image (center) was used by both models to generate the facade images on either side. Pix2pix-Unbalanced (left) introduces visual artifacts since, during training, the GAN loss term dominates the pixel-wise loss. Pix2pix-Unbalanced-AGB (right) can mitigate these artifacts by adaptively updating the $\beta$ term, which upper bounds the GAN loss by a factor of the pixel-wise loss gradients.
  }
    \label{fig:quali_pix2pix}
\end{figure}

\textbf{Regularization unit}
Each unit (Fig.~\ref{fig:dci}C) has three sequences consisting of 5x5 convolution, bias, and leakyReLU \cite{36_lrelu_xu2015empirical} layers. The output of the final iteration (Fig.~\ref{fig:dci}A) is (1) compared to the fully sampled reference image to generate a pixel-wise loss function, using MSE, and (2) paired with its corresponding zero-filled image and fed into the discriminator network to evaluate WGAN loss~\cite{29arjovsky2017wasserstein}. 

\subsubsection{Discriminator}
For our discriminator we use a convolutional ``PatchGAN'' \cite{18li2016precomputed}. The discriminator receives a pair of (1) $m_z$ and (2) $m_f$ or $G(K_u)$,  concatenated as two channels, and is able to penalize structure at the scale of image patches, from both channels. The architecture incorporates four convolutional layers with a stride of 2, each followed by batch normalization
and LeakyReLU. 
The last convolutional layer is flattened and then fed into a linear layer, for which each input value corresponds to a different patch in the input channels. The linear layer outputs a single value, used to calculate the discriminator's WGAN loss. {\color{black}Importantly, our discriminator receives pairs of zero-filled image and generate/real image. Paring those images allows the discriminator to match between features from the zero-filled image and generated/real image in the spatial space. In other words, such a discriminator can not be fooled by fake images that are inconsistent with their zero-filled image.}

\begin{figure}[t]
  \includegraphics[width=1.0\linewidth]{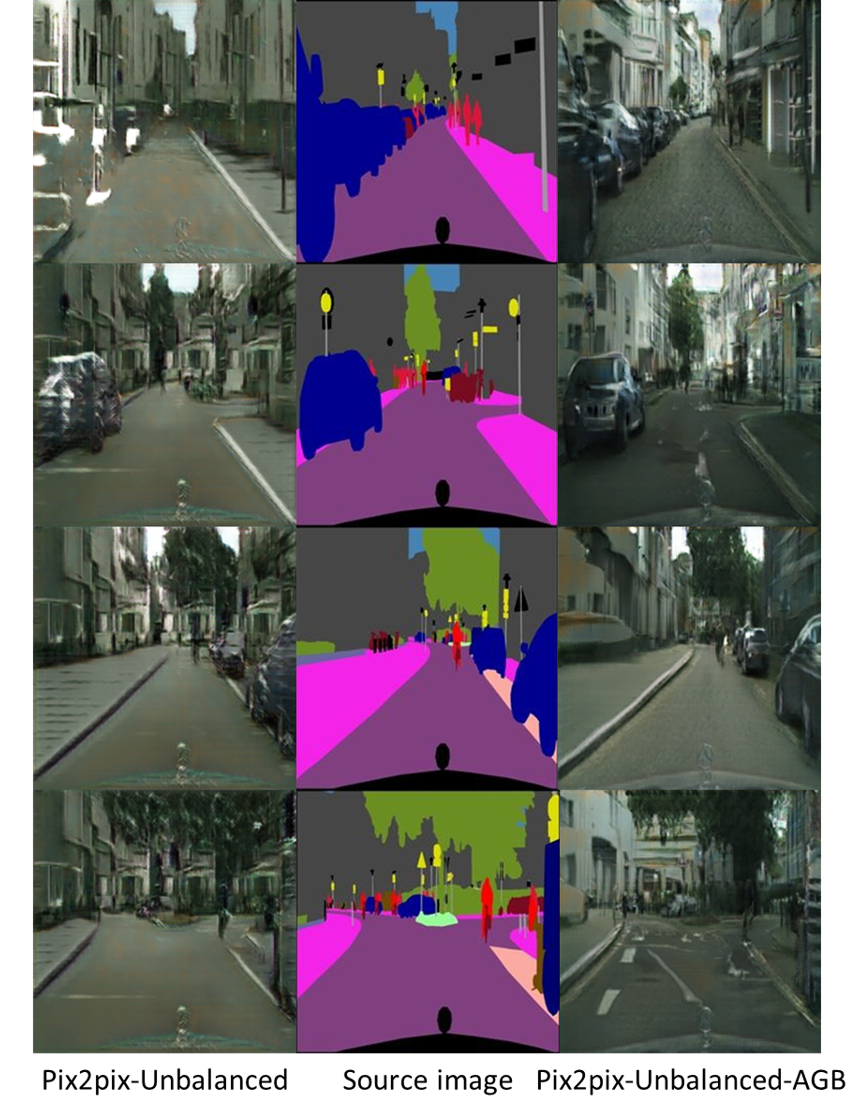}
  \centering
  \caption{{\color{black}Representative samples from the cityscape \cite{cityscapes} test set. The same annotated images (center) were used by both models to generate the urban street images on either side. Pix2pix-Unbalanced (left) introduces visual artifacts and fails to generate realistic car instances. Pix2pix-Unbalanced-AGB (right) yields higher-quality images.}
  }
    \label{fig:quali_cityscape}
\end{figure}

\begin{figure}[t]
  \includegraphics[width=0.85\linewidth]{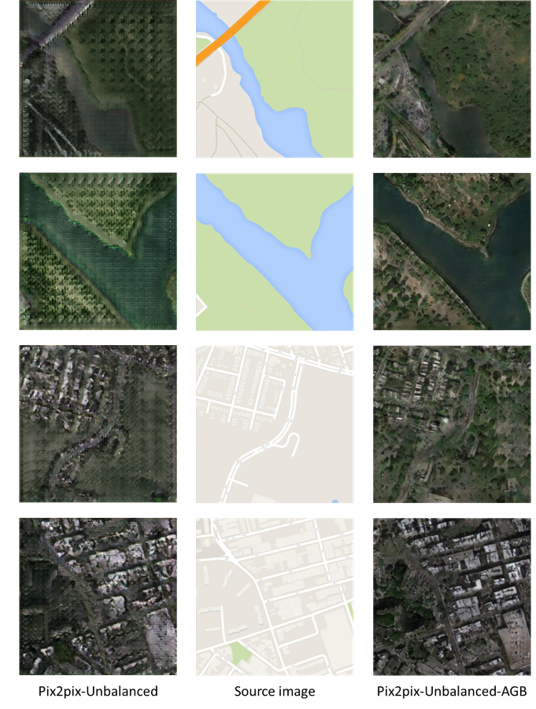}
  \centering
  \caption{{\color{black}Representative samples from the maps test set. The same map images (center) were used by both models to generate the aerial photographs on either side. Pix2pix-Unbalanced (left) introduces visual artifacts. Pix2pix-Unbalanced-AGB (right) yields higher-quality images.}
  }
    \label{fig:quali_maps}
\end{figure}

\begin{figure*}[t]
\centering
\begin{tabular}{ccc}
    \includegraphics[width=.3\linewidth]{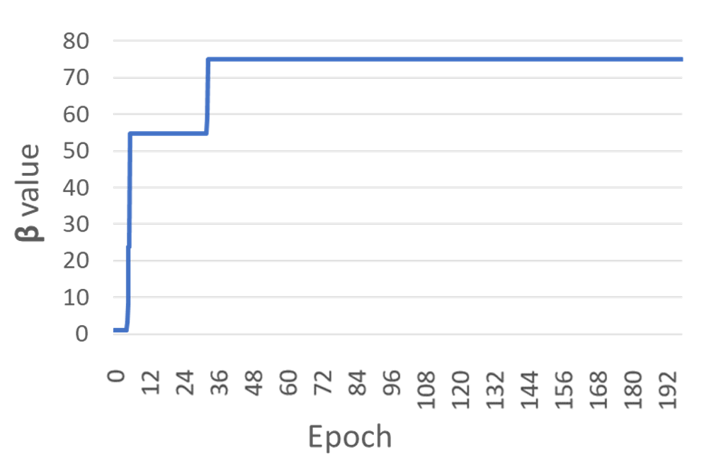}&
    \includegraphics[width=.3\linewidth]{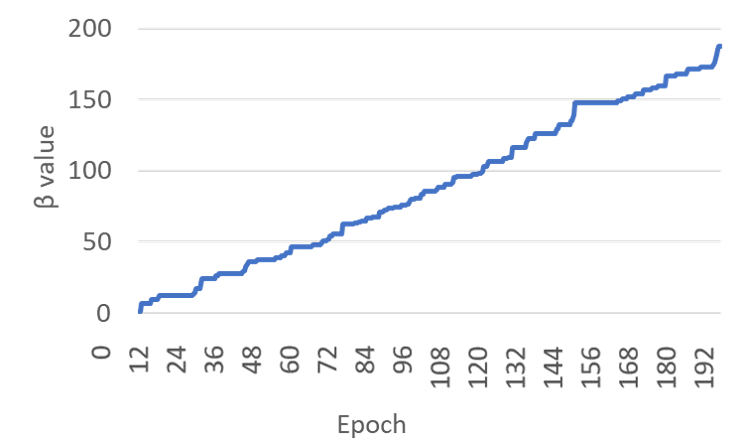}&
    \includegraphics[width=.3\linewidth]{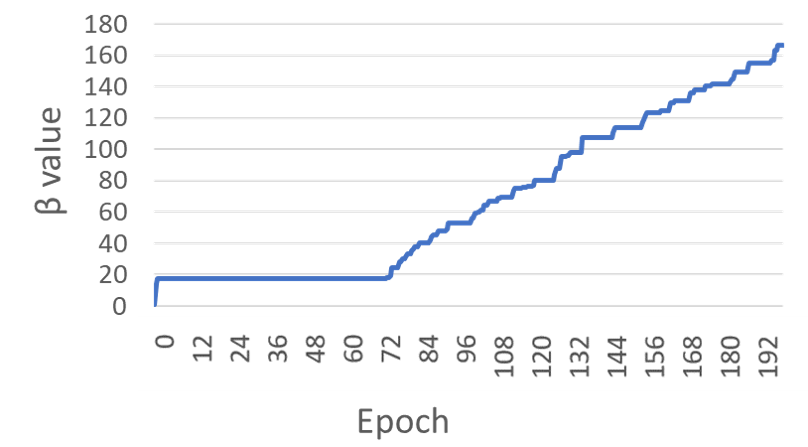}    \\
    (a) & (b) & (c) \\
    \end{tabular}
  \caption{{\color{black} $\beta$ value calculated versus epoch, for the Pix2pix-Unbalanced-AGB model trained to generate (a) facade images from annotations, (b) urban street scenes from annotations, and (c) aerial photographs from map images.}
    }
    \label{fig:beta_pix2pix}
\end{figure*}

\begin{figure*}[t]
\centering
\begin{tabular}{ccc}
    \includegraphics[width=.3\linewidth]{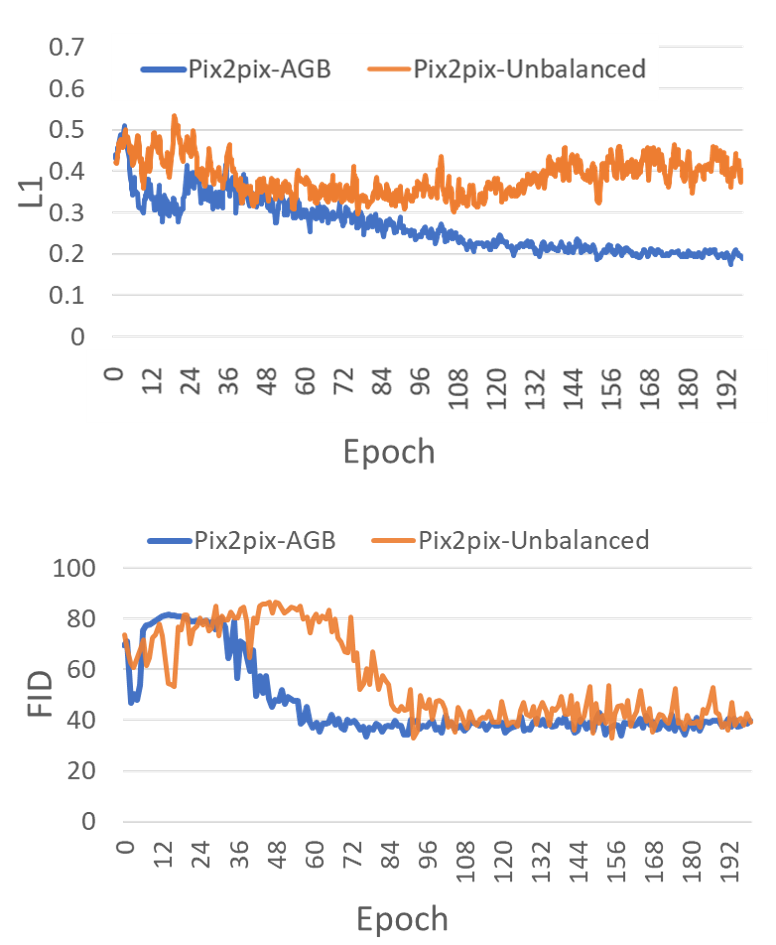}&
    \includegraphics[width=.3\linewidth]{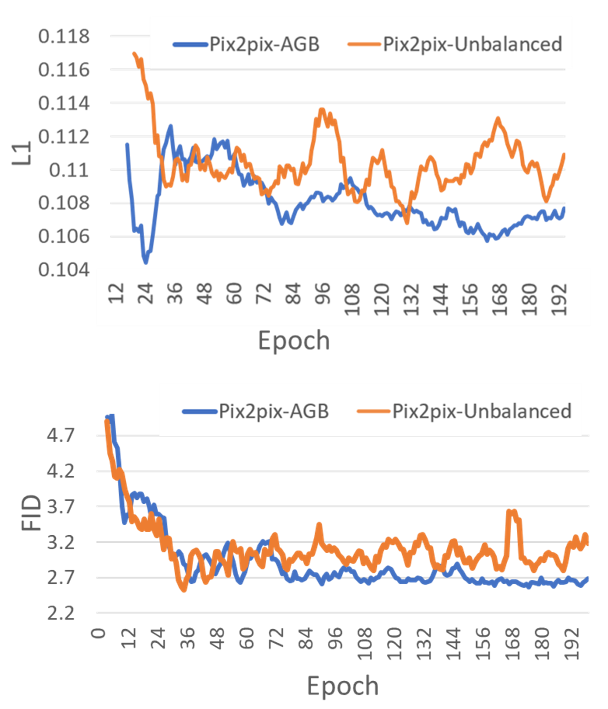}&
    \includegraphics[width=.3\linewidth]{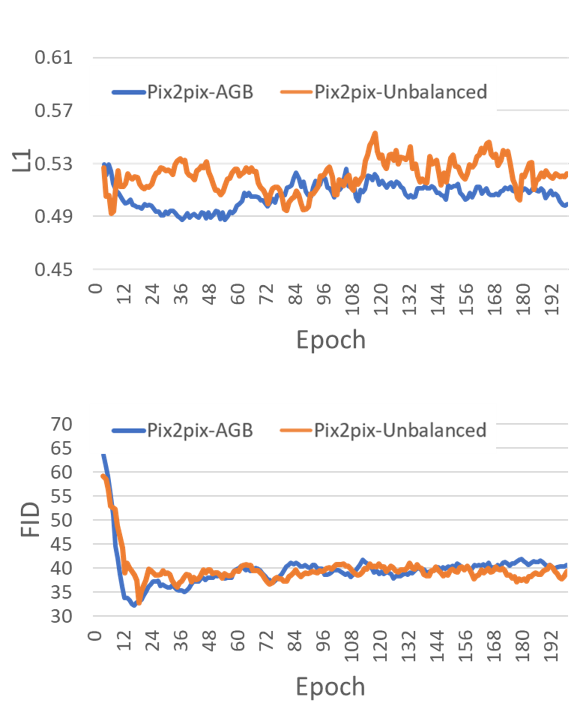}    \\
    (a) & (b) & (c) \\
    \end{tabular}
  \caption{{\color{black} L1 and FID per epoch, reported on the validation set, for both unbalanced models trained on (a) facade, (b) cityscapes, and (c) maps datasets. }
    }
    \label{fig:metrics}
\end{figure*}

\vspace{-8pt}
\section{Pix2Pix Experiments}

To show the generality of ABG beyond the MRI settings for which it was developed, we consider first the optimization of a multi-term adversarial loss function in the Pix2pix model \cite{20isola2017image}. This model transforms images from a source domain to a target domain. The Pix2pix objective incorporates a bi-term adversarial loss function, comprising a pixel-wise loss (L1) and a GAN loss. Pix2pix balances the two loss terms by means of weights crafted using a hyperparameter search. AGB does not require hyperparameter tuning. Therefore, it streamlines the development process by saving the exhaustive search required for carefully weighting multi-loss optimization terms.

The Pix2Pix experiments investigating the balancing of the GAN loss and the pixel-wise losses are conducted on three datasets: facade \cite{Tylecek13}, aerial-maps~\cite{20isola2017image} and cityscapes~\cite{cityscapes}. To this end, we trained two models for each of the three datasets. The first model of each dataset, utilizes the original Pix2pix scheme. The second employs the Pix2pix architecture with AGB training (denoted by Pix2pix-AGB). The two Facade models were trained to generate facade \cite{Tylecek13} images from facade annotations. The aerial-maps models were trained to generate aerial photographs from their matched map images. The cityscapes models were trained to generate urban street scenes from their matched semantic annotations.

In Fig.~\ref{fig:pix2pix_gradients}, we show moving averages, calculated on the norms of the GAN gradients, for the Pix2pix and Pix2pix-AGB models of all three experiments. In these experiments we do not employ early stopping, as suggested in~\cite{20isola2017image}. As can be seen, the gradients of Pix2pix are constantly increasing, while those of Pix2pix-AGB reamin bounded and range over a much smaller interval.

{\color{black}These experiments indicate that without applying an early stopping, Pix2pix could become unstable. Indeed, we verified that longer training (1000 epochs instead of the originally suggested 200 epochs) introduces visual artifacts.} On the other hand, the Pix2pix-AGB model, trained for 1000 epochs, yields images with higher quality (see Fig.~\ref{fig:Quali_pix2pix_1000}).

We further demonstrate the ability of AGB to improve convergence and mitigate artifacts in a Pix2pix model trained with a non-optimal weighting between its loss terms. {\color{black}To this end, we trained an additional two Pix2pix models, for each of the above three datasets,} while using non-optimal loss weighting. Hence, we multiply the Pix2pix GAN loss term by 100 (the original factor was 1), and keep the original $\lambda$ value (100) for the L1 loss. The first model {\color{black}trained for each dataset}, is a Pix2pix with conventional training, which minimizes the unbalanced objective above (denoted by Pix2pix-Unbalanced). The second is a Pix2pix model trained with AGB (denoted by Pix2pix-Unbalanced-AGB). Following the Pix2pix work \cite{20isola2017image}, all models were trained for 200 epochs.

Figure \ref{fig:quali_pix2pix} presents two representative images from the test set of the unbalanced facade models. It can be seen that Pix2pix-Unbalanced introduces visual artifacts, while Pix2pix-Unbalanced-AGB yields higher-quality images.

{\color{black}
Figure \ref{fig:quali_cityscape} presents representative images from the test set of the unbalanced cityscapes models. The Pix2pix-Unbalanced yields urban street images with visual artifacts, while the Pix2pix-Unbalanced-AGB, yields images with higher quality. Specifically, as can be seen in the figure, the Pix2pix-Unbalanced model seems to completely fail to generate car instances. For example, in the first row of the figure, the left side of the source image indicates a sequence of car instances. The corresponding generated image of the Pix2pix-Unbalanced model introduces visual artifacts in the cars' location, while the Pix2pix-Unbalanced-AGB model yields relatively-realistic car instances.

Figure~\ref{fig:quali_maps} presents representative samples from the test set of the unbalanced maps models. As can be seen, the Pix2pix-Unbalanced model introduces visual artifacts while the Pix2pix-Unbalanced-AGB yields substantially higher-quality images.}

Figure \ref{fig:beta_pix2pix} presents the $\beta$ value  {\color{black}versus epoch, for the above three Pix2pix-Unbalanced-AGB models}. As can be seen, in the Facade experiment, $\beta$ first updates to a value of $\sim$55, and then stabilizes at a value of $\sim$75. Perhaps unsurprisingly, the latter value is close to the non-optimal multiplication weight of the GAN loss (100). {\color{black} In the cityscape experiment, the $\beta$ value increases during training, up to a value of $\sim$190.
In the maps experiment, $\beta$ first updates to a value of $\sim$20, then $\beta$ starts to constantly increase up to a value of $\sim$165. Notably, the trends of all $\beta$ values are well correlated with the GAN gradients of the matched Pix2pix models (see Figure \ref{fig:pix2pix_gradients}). 
In the original Pix2pix work, the weighting was manually crafted after conducting a hyperparameter search.
}

Figure \ref{fig:metrics} exhibits the FID and L1 reported on a validation set, versus epoch, for the two unbalanced models {\color{black}of each dataset}. As can be seen in the figure, Pix2pix-Unbalanced-AGB converges faster, compared to Pix2pix-Unbalanced, yielding improved L1 score, along with better or similar FID performance.

\begin{table*}[t]
\begin{minipage}[c]{0.42\linewidth}
\caption{Comparison of our method with zero-filled images (ZF), and reconstruction using wavelets or TV~\cite{5lustig2007sparse}, PI~\cite{31ARC_beatty2007method}, {\color{black}VN \cite{8hammernik2018learning} and CNN-Cascade \cite{9schlemper2018deep}}. NMSE is w.r.t fully sampled image.}
\label{table:comprison-to-classic-methods}
\centering

\begin{tabular}{|l|c|c|}
\hline
Images & NMSE $\times 1000$  & {\color{black}FID} \\
\hline\hline
ZF & 115 &  {\color{black}173.0}\\
Wavelets & 18.7 &  {\color{black}138.4}\\
TV & 14.1 &  {\color{black}117.0}\\
PI & 18.9 &  {\color{black}109.0}\\
{\color{black}Variational Network} & 6.70 & {\color{black}23.3}\\ 
{\color{black}CNN-Cascade} & 7.22 & {\color{black}22.7}\\ 
cWGAN-AGB   & \textbf{3.39}  &  {\color{black}\textbf{18.7}}\\

\hline
\end{tabular}
\end{minipage}%
\hfill
\begin{minipage}[c]{0.56\linewidth}
\caption{{\color{black}Ablation analysis. First section: Ablation for DCI-Net generator, where all models were trained to minimize an MSE loss alone, with no GAN loss. Second section: Ablation on the GAN technique. All WGAN variants utilize best performing architecture from the first section, i.e. a DCI-Net with 20 iterations ("20I"), growth rate of G = 5 ("5G") and 40 kernels for each convolution ("40K").
}}  
\label{table:quanitative-results}
\centering
\begin{tabular}{|l|c|c|}
\hline
Experiment & NMSE $\times 1000$ & FID\\
\hline\hline
DCI-Net (5I-5G-160K) & 3.67 & 20.2 \\ 
DCI-Net (20I-1G-40K, no dense) & 3.46 & 19.3\\
DCI-Net (20I-5G-40K) & \underline{3.24} & 19.4\\
\hline
WGAN & 3.71  & 19.7 \\
cWGAN & 3.61 &  19.9\\
cWGAN-AGB (proposed) & \textbf{3.39} &  \textbf{18.7}\\
\hline
\end{tabular}
\end{minipage}
\smallskip
\smallskip
\caption{Mean of sharpness, SNR, contrast, artifacts and overall IQ scored for our proposed cWGAN-AGB, a baseline DCI-Net {\color{black}(which optimizes an MSE loss alone without any GAN)} and the fully-sampled images. Scores 1 to 5 indicate poor to excellent.
}
\label{table:blind-test}
\centering
\begin{tabular}{|l|c|c|c|c|c|c|}
\hline
Images & Sharpness & SNR & Contrast & Artifacts & Overall IQ\\
\hline\hline
Fully sampled & 5.0 & 3.3 & 4.0 & 4.0 & 4.5 \\
DCI-Net (20I-5G-40K) & 2.3 & 4.5 & 4.0 & 3.8 & 2.3 \\
cWGAN-AGB (proposed) & \textbf{3.8} & 3.8 & 4.0 & 3.8 & \textbf{3.5} \\
\hline
\end{tabular}
\end{table*}

\section{MRI Experiments}

\subsection{Dataset} 
Fully sampled brain MRI datasets (T1, T2, T1-FLAIR and T2-FLAIR in axial, coronal and sagittal orientations) were acquired with various k-space data sizes and various numbers of coils along with sensitivity maps estimated from separate calibration scans. In total, 2267 slices were acquired, of which 1901 were used to train the networks, 151 for validation and 215 for testing. In addition, during training, we also applied random horizontal flips and rotations (bounded to 20 degrees) to augment the training set. The data were retrospectively down-sampled using 12 central lines of k-space and a 1D variable-density sampling pattern outside the central region, resulting in a net under-sampling factor $R = 4$. As evaluation metrics, we compute both normalized mean square error (NMSE), and the Fr\'echet Inception Distance (FID)~\cite{30heusel2017gans}, which is a similarity measure between two datasets that correlates well with human judgment of visual quality and is most often used to evaluate the quality of images generated by GANs (see Sec.~\ref{FID_metric}).

\vspace{-8pt}
\subsection{Comparison with Baseline Methods}
We compare on the test set our cWGAN-AGB to CS methods that use wavelets or Total Variation (TV)~\cite{5lustig2007sparse} and to an autocalibrated PI method~\cite{31ARC_beatty2007method}. {\color{black}We also compare to the Cascade CNN \cite{9schlemper2018deep} and Variational Network (VN) \cite{8hammernik2018learning}, both trained using our same dataset and sampling pattern.} As can be seen in Table~\ref{table:comprison-to-classic-methods}, our proposed model produces significantly more accurate reconstructions than the other methods{\color{black}, as measured by both the NMSE and FID metrics}. 

\begin{figure*}[t]
  \includegraphics[width=0.95\linewidth]{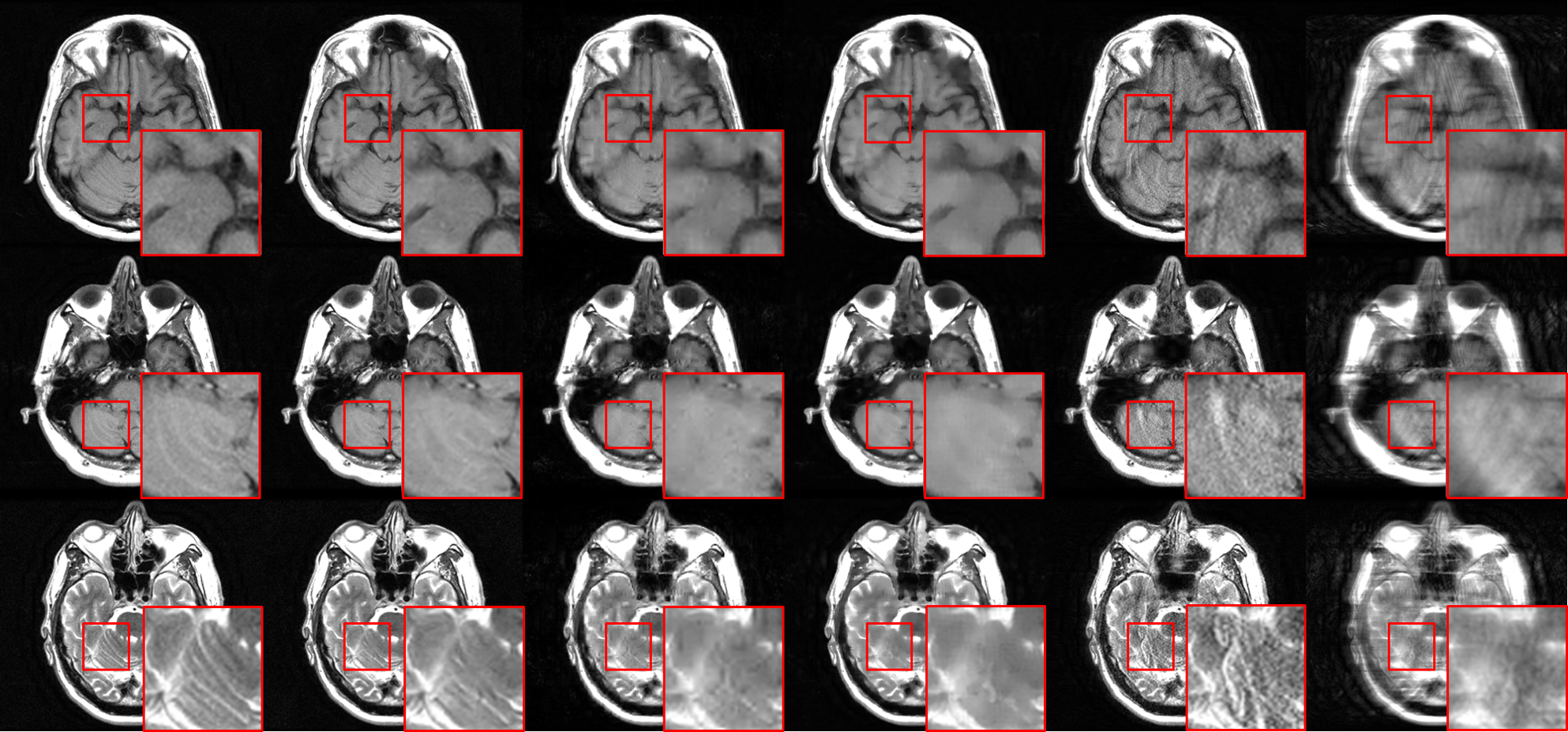}
  \centering
  \caption{{\color{black}Comparison with CS and PI methods. Left to right: fully sampled, cWGAN-AGB, wavelets, Total Variation, PI, zero-filled.}
  }
    \label{fig:qualitative-classic}
\end{figure*}

For the sake of completeness, we provide a qualitative comparison of our proposed model to compressed sensing methods using wavelets or TV ~\cite{5lustig2007sparse} and to PI ~\cite{31ARC_beatty2007method}, as shown in Fig.~\ref{fig:qualitative-classic}. It can be seen that our proposed method produces higher-quality images than traditional CS and PI methods, both in terms of perceptual quality and reconstruction error.

\vspace{-8pt}
\subsection{Comparing GANs Convergance}
\label{sec:convergance}
\begin{figure*}[t!]
\centering
\includegraphics[width=0.75\linewidth]{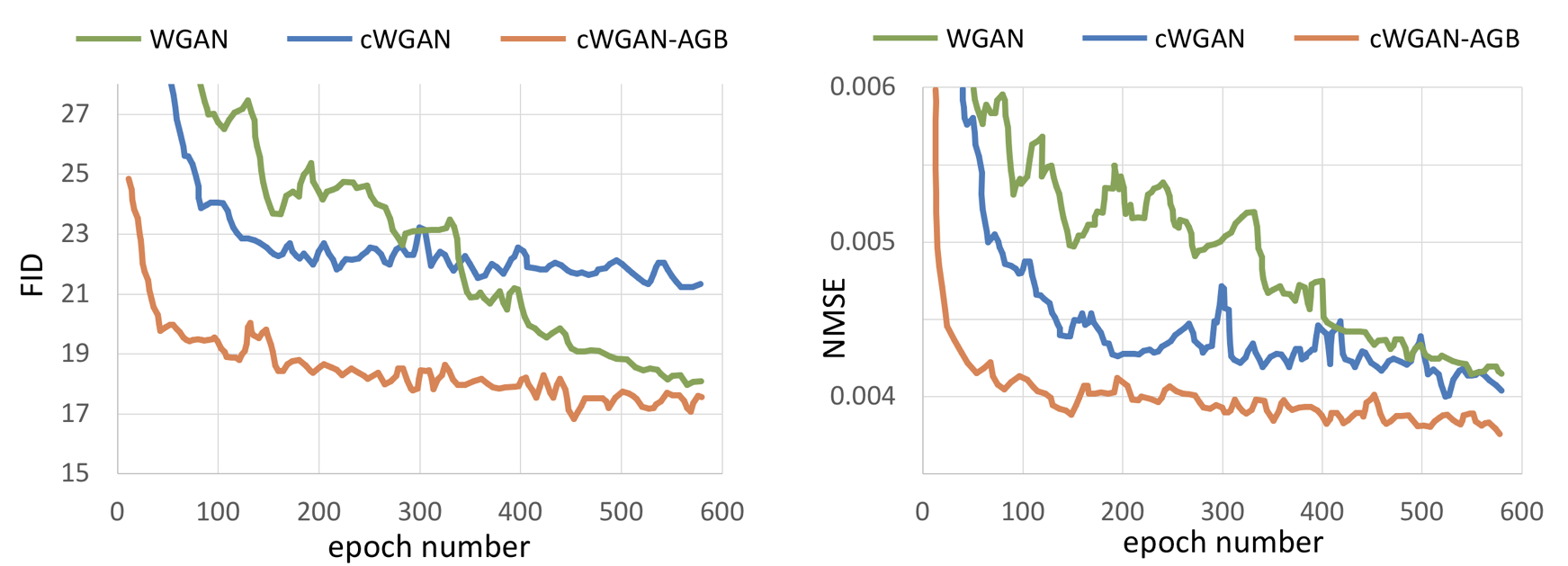}

\vspace{-8pt}
\caption{FID and NMSE during training, as evaluated on the validation set. Results are shown for WGAN, a vanilla cWGAN, and our cWGAN-AGB. 
}
\label{fig:val-curve}
\end{figure*}

To show the effectiveness of our method, we compared the convergence of our cWGAN-AGB model to that of cWGAN and WGAN, trained without AGB. During the training phase, FID and NMSE were evaluated on a hold-out validation set, for each epoch. Although WGAN suffers from a slow start (Fig.~\ref{fig:val-curve}), eventually it performs better on FID compared to cWGAN, but worse on NMSE (which can indicate a more realistic appearance at the cost of decreased fidelity). Our model converges better, with both scores decreasing substantially faster than the other techniques. {\color{black}For more experiments on AGB training, see the supplementary materials showing AGB applied to the Pix2pix model and evaluated on additional three different datasets.}

\vspace{-8pt}
{\color{black}
\subsection{Fr\'echet Inception Distance (FID)}
\label{FID_metric}
Fr\'echet Inception Distance (FID) \cite{30heusel2017gans} is a similarity measure between two datasets that correlates well with human judgment of visual quality and is most often used to evaluate the quality of images generated by GANs. We utilize FID as a quality metric to evaluate the similarity between the set of our generated images and the corresponding fully-sampled images. FID relies on the Fr\'echet distance calculated from two Gaussians each fitted on feature vectors taken from a pre-trained Inception network, one for the generated images and one for the fully-sampled images:
\begin{multline}
FID(P_{mf},P_{mg}) = \left\|\mu_{\text{p}}-\mu_{g}\right\|_{2}^{2}+ \\ \operatorname{Tr}\left(\Sigma_{p} + \Sigma_{g}-2\left(\Sigma_{p}\Sigma_{g}\right) ^ {\frac{1}{2}} \right)
\end{multline}
where $P_{mf}$ and $P_{mg}$ are sets of feature vectors extracted from an Inception network, for the fully sampled images and generated images, respectively. $\mu_{\text{p}}$, $\mu_{\text{g}}$ and $\Sigma_{p}$, $\Sigma_{g}$ are the mean and variance of the Gaussians fitted on $P_{m_f}$ and $P_{mg}$, respectively. 
}

\vspace{-8pt}
\subsection{Ablation Analysis}
\begin{figure*}[h]
\centering
\includegraphics[width=0.7\linewidth]{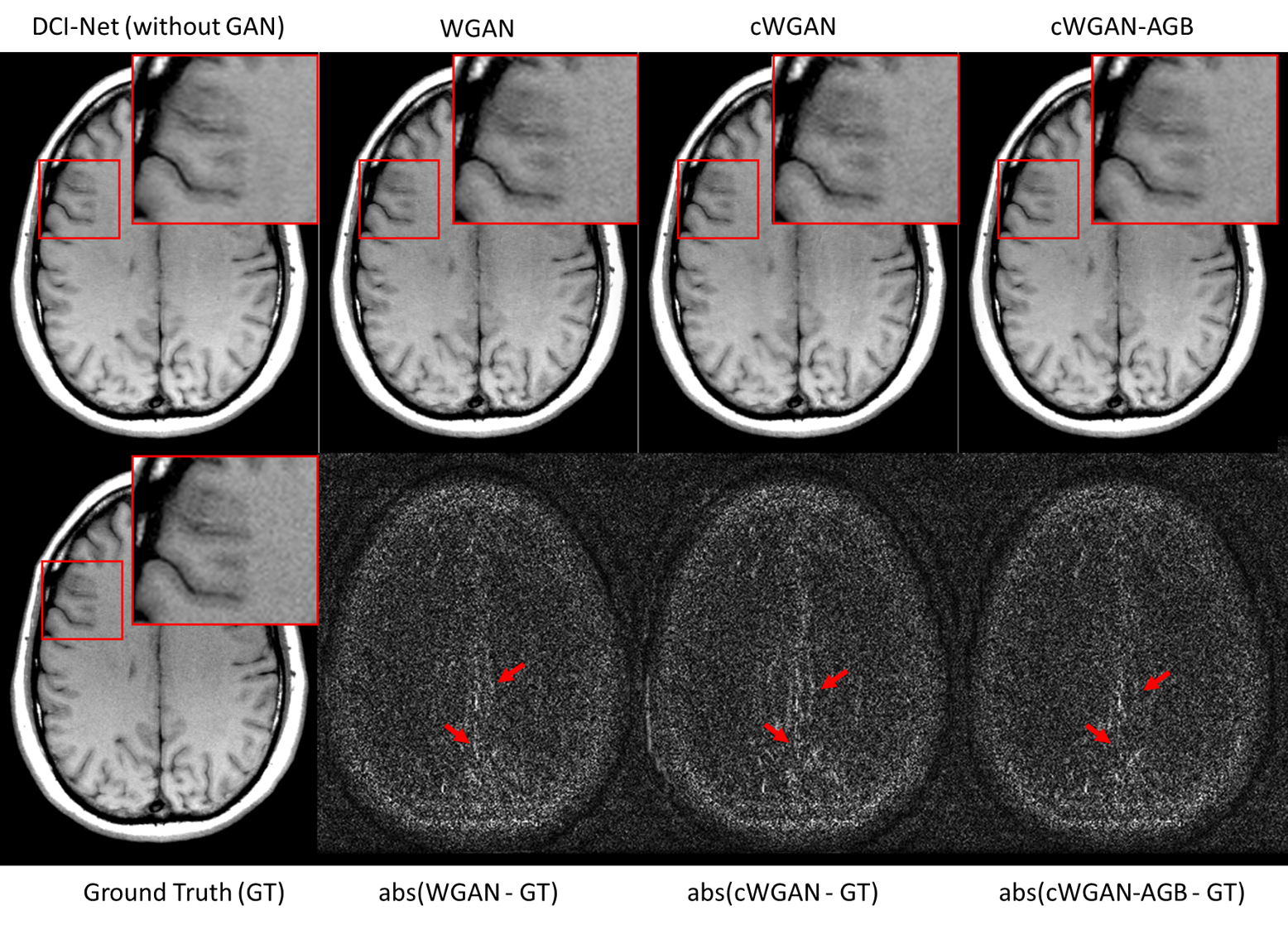}

\caption{A representative example with regions of interest showing the reconstruction of all models side-by-side (top row), along with the ground-truth fully-sampled image and the absolute difference images between ground-truth and each GAN based reconstruction (bottom row). cWGAN and cWGAN-AGB have better SNR than WGAN. cWGAN-AGB yields sharper images with fewer artifacts and more fine detail, while maintaining a more natural appearance. The baseline DCI-Net sometimes exhibits some blurring. 
}
\label{fig:imageCompare}
\end{figure*}

We compare, in Table~\ref{table:quanitative-results}, our cWGAN-AGB with three other models: 1) cWGAN, 2) WGAN, and 3) a baseline DCI-Net for undersampled MRI reconstruction, {\color{black}which optimizes an MSE loss alone without any GAN.} All models were evaluated with NMSE and FID on the test set. We found that (a) cWGAN and cWGAN-AGB have better SNR than WGAN, (b) cWGAN-AGB converges much faster than cWGAN or WGAN (Fig.~\ref{fig:val-curve}), has fewer artifacts, and performs better in both FID and NMSE measures (Table~\ref{table:quanitative-results}) and (c) although cWGAN-AGB has higher NMSE than the baseline DCI-Net, it performs better in FID and yields sharper images with more fine details while maintaining a natural image texture. {\color{black}A representative reconstruction can be seen in Fig.~\ref{fig:imageCompare}, where both WGAN and cWGAN models suffer from local inconsistencies with the ground truth image (red arrows). In the same area, our proposed method exhibits a more accurate reconstruction. In addition, Fig.~\ref{fig:imageCompare} shows a representative reconstruction from the baseline DCI-Net (trained without GAN loss), which exhibits some image blurring.} 

In Table~\ref{table:quanitative-results}, we also compare to baseline architectures, 
demonstrating the effectiveness of our key new architecture developments: (1) dense connections across all iterations, which strengthen feature propagation, making the network more robust, and (2) a relatively deep architecture of 20 iterations, comprising more than 60 convolutional layers, which brings increased capacity. {\color{black}We compared our generator to (1) an unrolled iterative network, similar to DCI-Net but without dense connections and (2) a 5-iteration DCI-Net with a similar number of learned parameters.} Employing dense connections significantly improved accuracy, and the deeper network produced 12\% lower mean NMSE than a shallower network with a similar number of learned parameters.

\vspace{-6pt}
\subsection{Visual Scoring}
To assess the perceptual quality of the resulting images we report a visual scoring conducted by four experienced MRI scientists. The same test set was ranked for cWGAN-AGB, the baseline DCI-Net and for the fully sampled images. The scoring was performed blindly and the images were randomly shuffled. The studies were taken from a cohort of seven healthy volunteers. Each study contained a full brain scan comprising 25-43 slices. For each study, image sharpness, signal-to-noise ratio (SNR), contrast, artifacts and overall image quality (IQ) were reported. {\color{black}Here images were rated on a scale of 1 to 5, where the numbers denote 1: not diagnostic, 2: limited, 3: diagnostic, 4: good, 5: excellent.} Table~\ref{table:blind-test} shows that cWGAN-AGB produced significantly sharper images than the baseline DCI-Net, at the cost of somewhat weaker denoising of the images.

\vspace{-8pt}
\subsection{Implementation Details}
\begin{figure}[t]
\centering
  \includegraphics[width=0.8\linewidth]{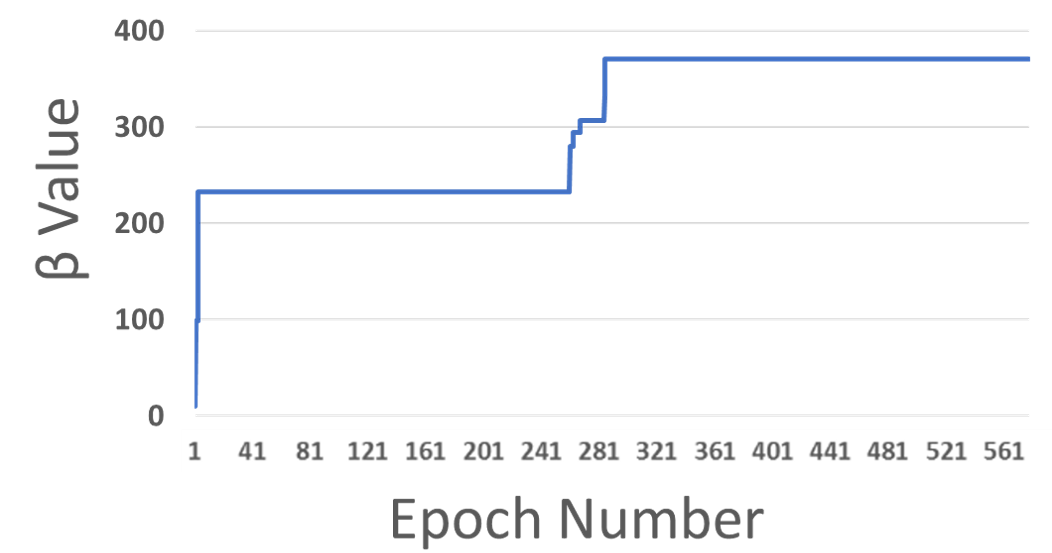}
  \caption{Beta value calculated per epoch, for cWGAN-AGB model. 
  }
    \label{fig:beta}
\end{figure}

Adam optimizer is used with a learning rate of $5\times10^{-4}$ for both generator and discriminator networks, with the momentum parameter $\beta_{m}$ = 0.9. Training is performed with TensorFlow interface on a GeForce GTX TITAN X GPU, 12GB RAM. For the proposed model with AGB training, $\beta$ is initialized to 10, without any hyperparameter exploration, and was found to increase in multiple steps during training to a value of 370 (see Fig.~\ref{fig:beta}). For the traditional GAN training, $\lambda$ is initialized to 100, after a hyperparameter search conducted on the values 10, 100, 1000. All models performed 600 epochs in $\sim$2 weeks of training, and the inference run time was 100ms per slice on a single GPU. Our code can be found at \url{https://github.com/ItzikMalkiel/AGB}.

{
\vspace{-8pt}
\section{Discussion of MRI Experiments}
The DCI-Net used as our generator is similar to [6] and [7] in that it is an unrolled optimization. While the CNN Cascade [7] alternates data-consistency and CNN blocks, the DCI-Net applies them in parallel within each iteration, in similar fashion to the VN [6]. However, while the VN learns, in addition to the convolutional filters, a set of nonlinear activation functions for each iteration, the DCI-Net employs leaky ReLU activations. But perhaps the biggest differences with these networks are our use of dense skip connections, combined with a relatively deep architecture.

One practical consideration is sensitivity to k-space undersampling pattern, and specifically, whether a network trained with one undersampling pattern and/or R value will face problems when inferencing with another pattern. While this could be addressed by training a separate network for each anticipated combination of undersampling pattern and R value, we have found that training a single network over a variety of patterns (including variable density and uniform) and R values (e.g. ranging from 2 to 6) can result in average NMSE values for test data that are within 6\% of those obtained from more focused networks.

While the upper part of Tab.~\ref{table:quanitative-results} shows that both the NSME and FID scores were improved for the deeper network, FID (unlike NMSE) did not likewise improve when dense connections were added. This can be attributed to (1) the ability of FID to indicate the perceptual quality of images, including blur level, and (2) the ability of  dense connections to improve accuracy, rather than mitigating blurring. 

While the methods described here were developed in the context of Cartesian k-space sampling, they could be readily extended to non-Cartesian trajectories such as interleaved spiral or radial. This would involve changing the generator's data-consistency block (Fig. 3D) to include an inverse gridding step (interpolating from the Cartesian grid onto the non-Cartesian trajectory) immediately following the 2DFFT into the k-space domain, and a second gridding step (interpolation onto the Cartesian grid) just prior to 2DIFFT into the image domain.

A number of investigators have searched for an optimal perceptual image-quality metric, but this has proven to be an elusive goal. While our metrics ranged from the relatively crude (NMSE), to a more sophisticated perceptually based score (FID), and supplemented by visual scoring using a 5-point-scale Mean Opinion Score (MOS), other metrics have been shown to have utility, including Structural Similarity Index (SSIM) \cite{wang2004image}, peak SNR, and Semantic Interpretability Score (SIS) \cite{seitzer2018adversarial}. SSIM has been widely adopted in medical imaging as a means of assessing quality based on degradation of structural information in the image. SIS has been used in cases where expert-provided segmentation labels are available, with Dice overlap calculated between ground-truth and network segmentations as a measure of visibility of segmented objects in the reconstructed images. And in \cite{schlemper2018stochastic}, a stochastic approach has been proposed to measure localized reconstruction uncertainty, by inferencing using a stochastic subset of sub-networks and measuring variance between different reconstructions. {\color{black}See supplementary for discussion about mitigating reconstruction instabilities.

\vspace{-8pt}
\subsection{DL-based MRI Reconstruction Instabilities}
A few components in our model are aimed to mitigate MRI reconstruction instabilities, similar to those discussed in \cite{pnas2020instabilities}, as well as potential hallucinations that may arise by the use of GANs. (1) we are the first to utilize a full conditional GAN architecture, in MRI reconstruction, which allows the discriminator to also verify fidelity, instead of solely reinforcing the realistic appearance of the images (as done in other papers that use GANs for MRI reconstruction). This property stems from the architectural choice of the discriminator, which receives both the generated image and the zero-filled image and therefore can match between the spatial features of both. (2) the unrolled architecture, alternating between convolutions and data consistency terms, reinforces the generator to produce images that are very close to the measure k-space. In our generator architecture, we apply a data consistency operation after every convolutional block. (3) the essence of the Adaptive Gradient Balancing is to ensure that the GAN component does not dominate the MSE loss. This entails a superiority to the MSE loss over the GAN loss since AGB decays the GAN gradients to be below a specific threshold defined by the MSE loss. Hence, AGB encourages the network to produce images that are more faithful to the images of full scans yet maintaining a natural image texture.
}}

\vspace{-8pt}
\section{Conclusions}

We present a novel undersampled MRI reconstruction model that employs a cWGAN 
with a novel multi-loss GAN training procedure. Our AGB training adaptively 
balances the adversarial and pixel-wise terms, streamlines the process of hyperparameter tuning (by saving the exhaustive search required for carefully weighting multi-loss adversarial terms), accelerates convergence and results in superior performance. By leveraging GANs to their fullest, the method generates sharper images with more fine detail and natural appearance than would otherwise be possible. In addition, dense connections are used to improve the performance of our unrolled iterative generator network. 
In the context of MRI reconstruction, a GAN based model can raise concerns about hallucination, where image details that do not appear in the ground truth are generated. We found that our method produces significantly less hallucination than other GANs. {AGB is demonstrated both as an MRI method and as a general computer vision method. AGB training could be beneficial for any model employing a multi-term adversarial objective, especially in the medical domain where there is considerable variability in the quality of the input and less experience in balancing GAN loss terms.}

\ifCLASSOPTIONcaptionsoff
  \newpage
\fi



%

\bibliographystyle{IEEEtran}
\bibliography{bibshort}

\ifpeerreview \else


{\color{black}
\vspace{-35pt}
\begin{IEEEbiography}[{\includegraphics[width=1in,height=1.25in,clip,keepaspectratio]{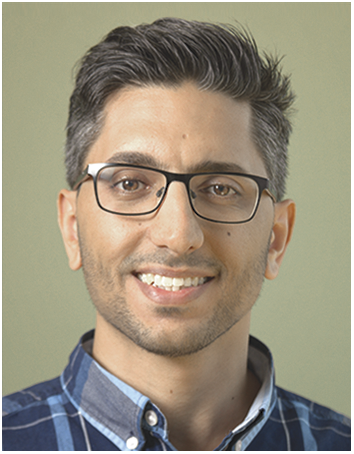}}]{Itzik Malkiel}
is a doctoral candidate in the School of Computer Science at Tel Aviv University, Israel, and a senior researcher on the Recommendation Machine Learning Research Team with Microsoft Corporation. He received his BSc degree in computer science from the Hebrew University, Israel, in 2010. He received his MSc degree in computer science from Tel Aviv University, Israel, in 2016. Prior to joining Microsoft, he was a research scientist with GE Global Research and a senior software engineer and project leader. His current research focuses on machine learning (ML) and deep learning and includes topics such as medical imaging, computer vision, natural language understanding, and ML applications in physics. Malkiel can be reached by email at itzik.malkiel@gmail.com.
\end{IEEEbiography}
\vspace{-35pt}
\begin{IEEEbiography}[{\includegraphics[width=1in,height=1.25in,clip,keepaspectratio]{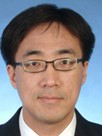}}]{Sangtae Ahn}
received the BS and MS degrees in mechanical engineering from KAIST (Korea Advanced Institute of Science and Technology), Korea, and the MS and PhD degrees in electrical engineering from the University of Michigan. He is a research scientist at GE Research. His research interests include image reconstruction and machine learning for medical imaging applications.
\end{IEEEbiography}
\vspace{-35pt}
\begin{IEEEbiography}[{\includegraphics[width=1in,height=1.25in,clip,keepaspectratio]{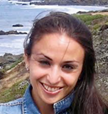}}]{Valentina Taviani}
received her PhD degree from the University of Cambridge, UK, in 2011. She worked as a research imaging scientist in the Radiological Sciences Laboratory at Stanford University before joining GE Healthcare in 2016, where she currently works as a senior applications engineer. Her research interests include deep learning-based image reconstruction, image enhancement and high-performance computing.
\end{IEEEbiography}
\vspace{-35pt}
\begin{IEEEbiography}[{\includegraphics[width=1in,height=1.25in,clip,keepaspectratio]{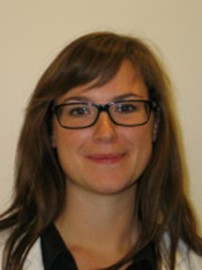}}]{Anne Menini}
received her PhD degree from the University of Lorraine, France in computer science in 2013. She worked as a research imaging scientist at GE Global Research in Munich, Germany before joining GE Healthcare in 2017, where she currently works as a lead scientist. Her research interests include advanced MRI reconstruction methods and motion correction technologies, and machine learning-based solutions for MRI.
\end{IEEEbiography}
\vspace{-35pt}
\begin{IEEEbiography}[{\includegraphics[width=1in,height=1.25in,clip,keepaspectratio]{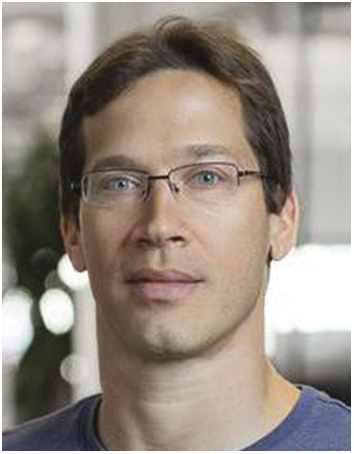}}]{Lior Wolf}
is a full professor in the School of Computer Science at Tel Aviv University, Israel, and a research scientist with Facebook AI Research. From 2003 to 2006, he conducted postdoctoral research at the Massachusetts Institute of Technology. In 2004, he received his PhD degree from the Hebrew University, Jerusalem. He is a European Research Council grantee. His awards include the International Conference on Computer Vision (ICCV) 2001 and ICCV 2019 honorable mention, and the best paper awards at the European Conference on Computer Vision 2000, the post-ICCV 2009 Workshop on eHeritage, the pre-Computer Vision and Pattern Recognition Workshop on Action Recognition, and the Internet Corporation for Assigned Names and Numbers 2016. His research focuses on computer vision and deep learning.
\end{IEEEbiography}
\vspace{-35pt}
\begin{IEEEbiography}[{\includegraphics[width=1in,height=1.25in,clip,keepaspectratio]{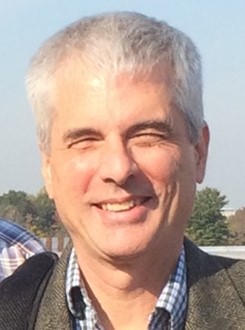}}]{Christopher Hardy}
received the AB degree in physics from Princeton University in 1977, and the PhD from the University of Illinois, Urbana-Champaign, in 1983. He is a senior principal scientist and Coolidge Fellow at GE Research. He is a fellow of the American Physical Society, the International Society for Magnetic Resonance in Medicine, and the American Institute for Medical and Biological Engineering. He was a visiting scientist at the University of Cambridge in 2012. His research interests include the application of deep learning to MR image reconstruction.
\end{IEEEbiography}
}




\fi

\end{document}